\documentstyle[aps,floats,amssymb,epsf]{revtex}

\font\lfrak=eufm10 at 10 pt

\newcommand{\lfr}[1]{\mbox{\lfrak #1}}

\def\beq{\begin{equation}}
\def\eeq{\end{equation}}
\def\beqa{\begin{eqnarray}}
\def\eeqa{\end{eqnarray}}
\def\hb{\hbar }

\def\vac0{{| 0 \rangle}}

\begin{document}
\draft
\title{SU(3) quasi-dynamical symmetry as an organizational mechanism
for generating nuclear rotational motions}
\author{C. Bahri and D.J. Rowe}
\address{Department of Physics, University of Toronto, \\
Toronto, Ontario M5S 1A7, Canada}
\date{\today}
\maketitle

\begin{abstract}
The phenomenological symplectic model with a Davidson potential is used
to construct rotational states for a rare-earth nucleus with microscopic
wave functions.
The energy levels and $E2$ transitions obtained are  in remarkably close
agreement (to within a few percent) with those of the rotor model with
vibrational shape fluctations that are adiabaticallly decoupled from the
rotational degrees of freedom.
An analysis of the states in terms of their SU(3) content shows that
SU(3) is a very poor dynamical symmetry but an excellent
quasi-dynamical symmetry for the model.
It is argued that such quasi-dynamical symmetry can be expected for any
Hamiltonian that reproduces the observed low-energy properties of a
well-deformed nucleus, whenever the latter are
well-described by the nuclear rotor model.
\end{abstract}




\section{Introduction}

A microscopic theory of nuclear structure would be very incomplete
without a satisfactory description of nuclear rotational states in terms
of many-nucleon quantum mechanics.
However, while the states of a truly rigid rotor can be handled with
ease \cite{Cas31}, they do not have square-integrable wave functions in
either a spherical vibrational-model or many-nucleon Hilbert space.
Moreover, the expansion of  liquid-like, soft-rotor, wave
functions on any spherical basis is slowly convergent.
This means that many major shells are required for a realistic
shell-model theory of nuclear rotational states.
It also means that a realistic calculation of nuclear rotational states in
terms of interacting nucleons, without a priori knowledge of the
kinds of correlations to expect, is an impossibly difficult task.
The fact remains that
nuclear rotational bands are exceedingly simple; they are essentially
characterized by a few intrinsic quadrupole moments and moments of
inertia. Furthermore, they are observed in a wide range of nuclei
throughout the periodic table. Thus, it would appear that the phenomenon
is remarkably robust and relies very little, for its existence,  on the
details of the two-nucleon interaction.

As demonstrated by calculations within the framework of the nuclear
symplectic model \cite{RosRow77,RosRow80,ParkAl84}, it is possible to
construct model rotational states with many-nucleon wave functions.
The problem is to understand why the model works as well as
it does; i.e, why the many residual interactions that strongly
break symplectic symmetry, by mixing states of different irreps, do not
destroy the predictions of the model.

To understand this, the
effects of the dominant symmetry breaking interactions  have been
explored one at a time.
It has been shown that, while a spin-orbit interaction may mix SU(3)
 irreps strongly, it does so, in large-dimensional irreps, in a highly
coherent way such that the rotational spectrum of the model survives
\cite{Roc88}.
Similar coherent mixings of SU(3) irreps by pairing forces
were found while investigating the transition of a many-fermion model
\cite{RowBW}
from a superconducting phase to a rotational phase with variation of the
relative strengths of short-range (pairing) and long-range
(quadrupole-quadrupole) interactions \cite{BahRowWij98}.
In this paper we examine in some detail the very
strong mixing of major harmonic-oscillator shells and SU(3) irreps, within
the framework of the symplectic model.
Again the mixings are extraordinarily coherent.
We refer to this coherent mixing as a {\it quasi-dynamical symmetry\/}
\cite{BahRowWij98,NAC}.

We conjecture that, because the
residual interactions separately preserve quasi-dynamical symmetry, they
continue to do so when combined for suitable ranges of their strengths.
This conjecture is strongly supported by the
observation of rotational states in nuclei which demonstrates
unequivocally that they do survive.

The discovery of quasi-dynmical symmetry gives optimism that
realistic calculations of rotational states, which simultaneously
take into account the mixings between major shells and, for example, the
effects of spin-orbit and short-range (e.g., pairing)
interactions,  may be possible.
Up to the present time such calculation have not been possible except
within the framework of a highly constrained (e.g., Hartree-Bogolyubov)
approximation.

The calculations reported here are carried out with a simple Davidson
interaction potential \cite{Dav32}.
This interaction preserves a higher, symplectic model,
symmetry which makes calculations in a very large multi-shell space
possible.
With this interaction, the Hamiltonian for a
diatomic molecule was recently diagonalized by a simple non-linear
transformation from a spherical vibrator basis to a soft rotor basis
using an
$\lfr{su}(1,1)$  spectrum generating algebra (isomorphic to $\lfr{
sp}(1,{\Bbb R})$) \cite{RowBah98}.
 A similar transformation \cite{EllPar86,RowBah98} gives rotational
states in the Bohr-Mottelson collective model \cite{BohMot52}.

For a many-nucleon nucleus, one can construct a Hamiltonian comprising
the many-nucleon kinetic energy, a spherical shell model potential, and
the same Davidson potential as used to obtain rotational states in
the pheomenological Bohr-Mottelson model.
The Davidson potential is a scalar function of the nuclear quadrupole
moments expressed in terms of nucleon coordinates.
Thus, it is microscopic and rotationally-invariant.
More importantly, it is expressible in terms of an $\lfr{sp}(3,{\Bbb
R})$ spectrum generating algebra and and gives a Hamiltonian that is
diagonalizable within a single irrep of the symplectic model
\cite{RosRow77,RosRow80,ParkAl84}. We are not able to give analytical
expressions for its wave functions, as we were \cite{RowBah98} for the
diatomic molecule and Bohr-Mottelson model.
But, we are able to compute its states numerically and expand
them on a spherical shell-model basis. This is done by diagonalizing the
Hamiltonian in a basis that reduces the dynamical subalgebra chain
\beq \lfr{sp}(3,{\Bbb R}) \supset \lfr{su}(3) \supset \lfr{so}(3)
\, ,\eeq
where $\lfr{su}(3)$ is the symmetry algebra of the spherical harmonic
oscillator shell model \cite{Ell58}.

Symplectic model calculations in an $\lfr{su}(3) \supset \lfr{so}(3)$
basis have been developed to a fine art using VCS
(vector coherent state)
methods \cite{Row84,Row85} to calculate matrix elements.
Furthermore, they can be carried out in large spaces with relative ease.

An important property of the $\lfr{sp}(3,{\Bbb R})$ algebra is that it
contains, as a subalgebra, a spectrum generating algebra for a rigid-rotor
model, namely Ui's $[{\Bbb R}^5]\lfr{so}(3)$ algebra \cite{Ui}.
However, one cannot diagonalize a symplectic model Hamiltonian in a basis
that reduces the subalgebra chain
\beq \lfr{sp}(3,{\Bbb R}) \supset [{\Bbb R}^5]\lfr{so}(3) \supset
\lfr{so}(3)\, ,\eeq
because, as mentioned above, rigid-rotor wave functions are not
expandable in a shell-model basis (in technical terms, the decomposition
is not a direct sum;  it is a direct integral.).
 One notes, however, that
the Hamiltonian we are considering as a model of a soft rotor has two
components:  a spherical shell model part which reduces  the
$\lfr{sp}(3,{\Bbb R})\supset \lfr{su}(3)\supset \lfr{so}(3)$ chain
and a Davidson potential which reduces
$\lfr{sp}(3,{\Bbb R}) \supset [{\Bbb R}^5]\lfr{so}(3) \supset
\lfr{so}(3)$.
Thus, we expect results that lie between the two limits.
In fact, the results of our calculations prove to be such that both
$[{\Bbb R}^5]\lfr{so}(3)$ and  $\lfr{su}(3)$ are extremely good
quasi-dynamical symmetries.
This is a  remarkable result, because although
$\lfr{su}(3)$ and the rigid-rotor algebra have similar algebraic
structures, they have very different physical interpretations.

A  Lie algebra $\lfr{g}$ is said to be a quasi-dynamical
symmetry for a Hamiltonian $H$ if  the matrix elements of $\lfr{g}$
between a set of eigenstates of $H$ are equal to those of an irrep of
$\lfr{g}$, even though the eigenstates in question do not belong to a
subspace of the Hilbert space for $H$ that is irreducible under the
action of $\lfr{g}$.
At first sight this would be appear to be an unlikely
physical situation.
Indeed, if matrix elements of an algebra $\lfr{g}$ were found in some
situation to be equal to those of an irrep, one
might be tempted to infer that $\lfr{g}$ is a full dynamical symmetry
for the Hamiltonian.
The soft rotor model is the
prototype of a situation
where this is not the case, as we now show.

Let $\{ \Phi_{KLM}(\beta,\gamma)\}$ denote a basis of wave functions for a
rigid rotor with intrinsic quadrupole moments
\beq \bar Q_0 = k \beta\cos\gamma \, ,\quad
\bar Q_1=\bar Q_{-1}=0\,, \quad \bar Q_2 = \bar Q_{-2} = {1\over
\sqrt{2}}k\beta\sin\gamma\,,\eeq
where $k$ is a suitable constant, cf.\ Eq.\ (\ref{eq:betagamma}).
Then soft-rotor wave functions can be expressed
\beq \Psi_{\alpha KLM} = \int f_\alpha (\beta,\gamma)\,
\Phi_{KLM}(\beta,\gamma)\,  {\rm d}v\, ,\eeq
where d$v$ is a suitable volume element for $\beta$ and $\gamma$.
One finds that the matrix elements between all states of the same
$\alpha$ are equal to those of a rigid rotor of deformation
\beq \langle \bar Q_0\rangle = \int  k \beta\cos\gamma\, f_\alpha
(\beta,\gamma)
\,  {\rm d}v \, ,\quad
\langle \bar Q_2\rangle =\int {1\over\sqrt{2}}k\beta\sin\gamma\, f_\alpha
(\beta,\gamma)
\,  {\rm d}v \, . \eeq
Thus, the states of a given $\alpha$ span an {\it embedded
representation\/} of the rigid-rotor algebra in the terminology of ref.\
\cite{RowRoc88} and we say that the rigid-rotor algebra is a
quasi-dynamical symmetry for the soft rotor.

An examination of the physics of the situation quickly reveals that the
rigid-rotor algebra is a quasi-dynamical symmetry whenever the rotations
are adiabatic relative to the complementary vibrational motions.
Thus, one obtains the rigid-rotor algebra as a quasi-dynamical symmetry
for a rotor-vibrator Hamiltonian whenever the rotational-vibrational
(i.e., Coriolis and centrifugal) coupling interactions are
negligible or omitted. They can be taken into account subsequently, as
quasi-dynamical symmetry-breaking perturbations, as is standard in
the pheomenological rotor model.

Now it is known \cite{C84} that the $\lfr{su}(3)$ algebra
contracts to the rigid-rotor algebra  $[{\Bbb R}^5]\lfr{so}(3)$ and
that this contraction is realized for large-dimensional irreps of
$\lfr{su}(3)$. Thus, we say that $\lfr{su}(3)$ is a quasi-dynamical
symmetry for a sequence of states if their expansions on an $\lfr{su}(3)$
basis are of the form
\beq  \Psi_{\alpha KLM} = \sum_{\lambda\mu} C_{\alpha\lambda\mu}
\Phi^{\lambda\mu}_{KLM} \, ,\eeq
where $\{\Phi^{\lambda\mu}_{KLM}\}$ is a basis of states of a
large-dimensional
$\lfr{su}(3)$ irrep of highest weight $(\lambda,\mu)$. Note that
shell-model states can always be expanded on an  $\lfr{su}(3)$ basis.
However, we only obtain  $\lfr{su}(3)$ as a quasi-dynamical symmetry if
the coefficients $\{ C_{\alpha\lambda\mu}\}$ are independent of $KLM$ for
a useful range of values of the latter.

The fact that $\lfr{su}(3)$ is a good
quasi-dynamical symmetry for major-shell mixing Hamiltonians is extremely
important for achieving the eventual goal of including $\lfr{sp}(3,{\Bbb
R})$ symmetry-breaking interactions in the symplectic model.
For although, interactions like the spin-orbit and pairing interactions
break $\lfr{sp}(3,{\Bbb R})$, they have been shown to preserve
quasi-dynamical $\lfr{su}(3)$ symmetry \cite{Roc88,BahRowWij98}.
This is  expected to occur, to a good approximation, whenever the
rotational motions one is describing are known from experiment to be
adiabatic.

The concept of quasi-dynamical symmetry can be regarded simply as
a group-theoretical expression of the
standard methods for handling adiabatic decoupling of collective motions
along the lines of the Born-Oppenheimer approximation \cite{BornO} and the
standard nuclear rotor model. A formulation in the precise language of
group theory has the advantage that it opens the concept up to more
general application. An overview of the concept was given in a recent
conference report \cite{NAC}.

In Sect.\ \ref{sect:Hmodel}, the Hamiltonian of the nuclear symplectic
model is described and the physical motivation behind the model in
connection with Bohr-Mottelson collective model \cite{BohMot52} is
explained.
Sect.\ \ref{sect:algebras} defines the  $\lfr{sp}(3,{\Bbb R})$ and its
subalgebras relevant to this analysis.
 Sect.\ \ref{sect:irreps} describes the construction of basis states
and matrix elements within the three dimensional harmonic oscillator
space.
The results are given in Sect.\ \ref{sect:results} and some
conclusions are
drawn in Sect.\ \ref{sect:conclusions}

\section{The Symplectic Model Hamiltonian}

\label{sect:Hmodel}
The Hamiltonian of the nuclear symplectic model, in its simplest
form,  consists of two parts:
\begin{equation}
H=H_{0}+V(Q) \, ;
\label{eq:Hsp3r}
\end{equation}
a three dimensional many-particle
harmonic oscillator (shell-model) Hamiltonian
\begin{equation}
H_{0}=\sum_{n=1}^{A}\left( {\frac{p_{n}^{2}}{2m}}+{\frac{1}{2}}m\omega
^{2}x_n^2\right)
\end{equation}
and a collective potential
$V(Q)$, where $Q$ is the quadrupole tensor for the nucleus.

 The collective potential $V(Q)$ is
a rotationally-invariant function of the quadrupole moments.
In a Cartesian basis, the quadrupole moments for a nucleus
are given  by
\begin{equation}
Q_{ij}={\Bbb Q}_{ij}-\textstyle\frac{1}{3}\delta _{ij}\sum_{k=1}^{3}
{\Bbb Q}_{kk}
\, ,
\label{eq:quadrupole}
\end{equation}
where ${\Bbb Q}$ is a monopole-quadrupole tensor with components
\begin{eqnarray}
{\Bbb Q}_{ij} &=&\sum_{n=1}^{A}(x_{ni}-X_{i})(x_{nj}-X_{j})  \nonumber \\
&=&\sum_{n=1}^{A}x_{ni}x_{nj}-\frac{1}{A}\sum_{m,n=1}^{A}x_{mi}x_{nj} \, ,
\label{eq:monopole-quadrupole}
\end{eqnarray}
and $\{X_i\}$ are the components of the center-of-mass vector
${\bf X} = {1\over A} \sum_n {\bf x}_n$.
Removal of the center-of-mass contribution to the quadrupole moments in
this way, ensures that the spurious $N$-phonon center-of-mass states
of the Hamiltonian remain, unmixed with other states, at an excitation
energy of $N\hbar\omega$.

\begin{figure}[tbp]
\epsfxsize=4in
\centerline{\epsfbox{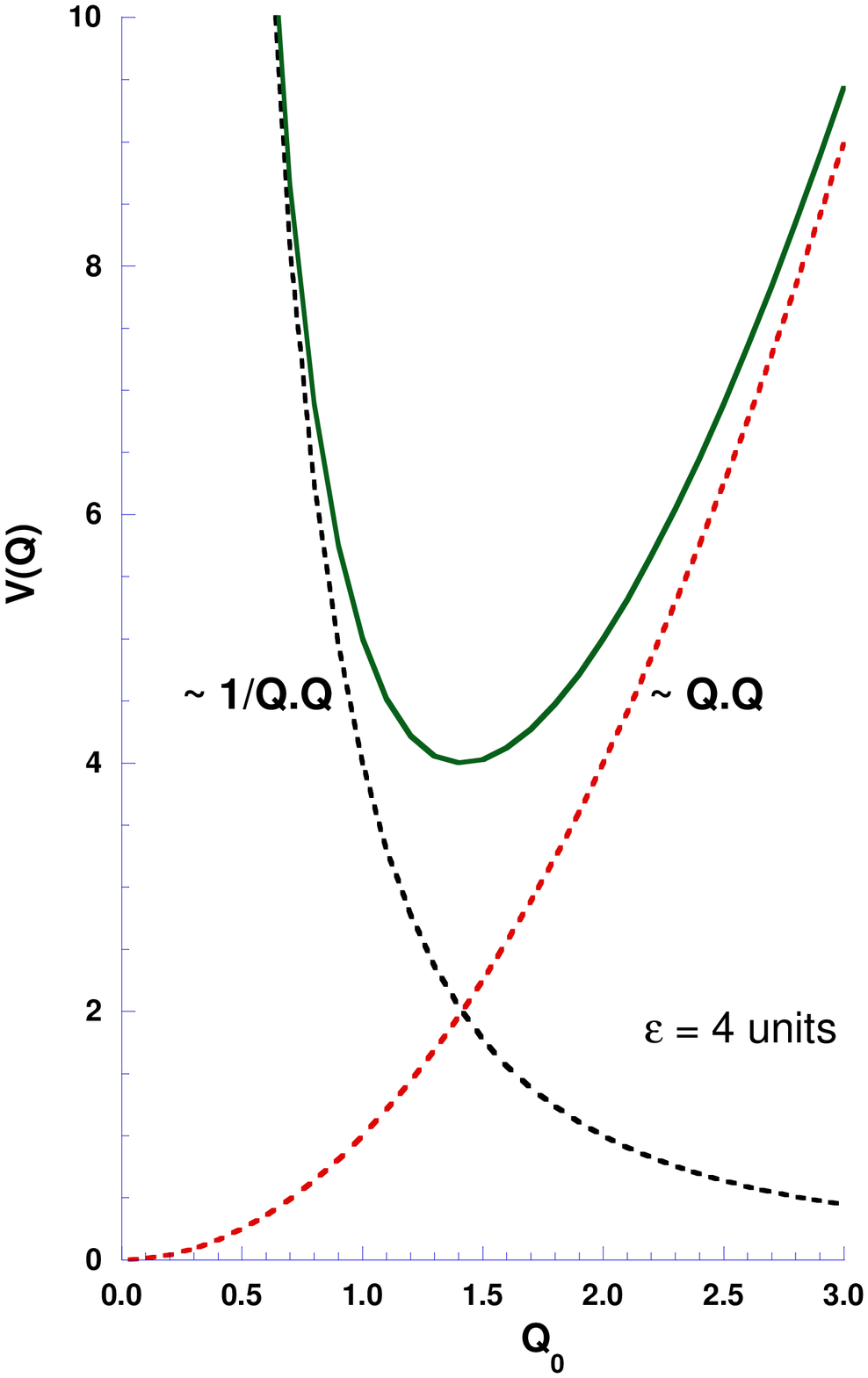}}
\caption{The Davidson potential; its two
components are shown separately.}
\label{fig:Vshape}
\end{figure}

A general, rotationally-invariant potential can be expressed
\begin{equation}
V(Q)=V(Q\cdot Q,Q\cdot Q\times Q)\,,  \label{eq:V_Q2Q3}
\end{equation}
as a function of the quadratic and cubic scalars
\begin{eqnarray}
Q\cdot Q &=&\sum_{\mu =-2}^{2}(-1)^{\mu }Q_{-\mu }Q_{\mu }
\;\propto {\rm Tr} Q^2 \,,  \nonumber \\
Q\cdot Q\times Q &=&\sum_{\mu ,\nu ,\rho =-2}^{2}(-1)^{\mu }(2\nu ;2\rho
|2\mu )Q_{-\mu }Q_{\nu }Q_{\rho }
\;\propto {\rm Tr} Q^3\,,  \label{eq:Q2Q3}
\end{eqnarray}
where
\[
Q_{\mu }=\sum\nolimits_{n}r_{n}^{2}Y_{2\mu }(\hat{\bf {r}}_{n})
\]
with $r_{n}=\left| {\bf x}_{n}-{\bf X}\right| $ and $\hat{\bf
{r}}_{n}= ({\bf x}_{n}-{\bf X})/r_n $.
In the study reported in this article,
the collective potential is taken to be the Davidson
potential
\begin{equation}
V(Q)=\chi \left( Q\cdot Q+\frac{\varepsilon }{Q\cdot Q}\right) \,.
\label{eq:V_Q}
\end{equation}
This potential is shown as a function of $\bar Q_0 = \sqrt{Q\cdot Q}$ in
Fig. \ref {fig:Vshape}.

The value of the parameter $\varepsilon$ determines the value
 of $\bar Q_0$ at which the potential has its minimum
value.
Thus, $\varepsilon$ is chosen such the potential has a minimum at the
observed deformation of the nucleus under investigation.
The strength $\chi$ of
the potential is then  set such that the lowest energy $(L=0)$ wave
function that emerges is such that the expectation value ${\langle
Q\cdot Q\rangle}$ in this state is equal to the value of $\bar Q_0^2$ at
which the potential is a minimum.  We refer to this as the self-consistent
value of
$\chi$.

\section{The symplectic algebra and its subalgebras}\label{sect:algebras}

The quadrupole moments $\{Q_{\mu }\}$, the many-nucleon kinetic energy $%
\sum_{n}p_{n}^{2}/2m$, and the harmonic oscillator potentional ${\frac{1}{2}}%
m\omega ^{2}\sum_{n}r_{n}^{2}$ are all elements of an $\lfr{sp}(3,{\Bbb
R})$ Lie algebra.
Thus, $\lfr{sp}(3,{\Bbb R})$ is a spectrum generating algebra for $H$;
i.e., it is the Lie algebra of a dynamical group for $H$.
It folows that the eigenstates of $H$ belong to a single
irreducible representation of Sp$(3,{\Bbb R})$.

The complex extension  $\lfr{sp}_{{\Bbb C}}(3,{\Bbb R})$ of
$\lfr{sp}(3,{\Bbb R})$ is spanned by the operators (in a Cartesian
coordinate system)
\begin{eqnarray}
A_{ij} &=&\sum_{n=1}^{A}b_{ni}^{\dagger }b_{nj}^{\dagger }-\frac{1}{A}%
\sum_{m,n=1}^{A}b_{mi}^{\dagger }b_{nj}^{\dagger }\,,  \nonumber \\
B_{ij} &=&\sum_{n=1}^{A}b_{ni}b_{nj}-\frac{1}{A}\sum_{m,n=1}^{A}b_{mi}b_{nj}%
\,,  \label{eq:sp3Rcomp} \\
C_{ij} &=&\frac{1}{2}\sum_{n=1}^{A}(b_{ni}^{\dagger
}b_{nj}+b_{nj}b_{ni}^{\dagger })-\frac{1}{2A}\sum_{m,n=1}^{A}(b_{mi}^{%
\dagger }b_{nj}+b_{nj}b_{mi}^{\dagger })\,,  \nonumber
\end{eqnarray}
where $b^\dagger_{ni}$ and $b_{ni}$ are the dimensionless harmonic
oscillator raising and lowering (boson) operators
\begin{eqnarray*}
b_{ni}^{\dagger } &=&\frac{1}{b_{0}\sqrt{2}}\left( x_{ni}-\frac{i}{m\omega }%
p_{ni}\right) \,,\\
b_{ni} &=&\frac{1}{b_{0}\sqrt{2}}\left( x_{ni}+\frac{i}{m\omega }
p_{ni}\right)\,,
\end{eqnarray*}
in units of the oscillator length $b_{0}=(\hbar /m\omega )^{1/2}$.
The latter operators satisfy the commutation relations
\begin{eqnarray}
{\lbrack b_{mi},b_{nj}^{\dagger }]} &=&\delta _{mn}\delta _{ij}
{\Bbb I}\,,  \nonumber
\\
{\lbrack b_{ni},{\Bbb I}]} &=&{[b_{ni}^{\dagger },{\Bbb I}]}=0\,, \\
{\lbrack b_{mi},b_{nj}]} &=&{[b_{mi}^{\dagger },b_{nj}^{\dagger }]}=0\,,
\nonumber
\end{eqnarray}
of a Heisenberg-Weyl algebra.

The symplectic algebra $\lfr{sp}(3,{\Bbb R})$ contains many subalgebras,
including the  $\lfr{u}(3) \supset  \lfr{su}(3)$ chain of Elliott's SU(3)
model \cite{Ell58} and the   $\lfr{rot}(3)$ $(= [{\Bbb
R}^{5}]\lfr{so}(3))$ rigid-rotor algebra of Ui's model \cite{Ui}.
The $\lfr{u}(3)$ subalgebra is spanned by the $\{ C_{ij}\}$
operators.
One sees, for example, that $\lfr{u}(3)$ contains the harmonic oscillator
Hamiltonian
$H_{0}=\hbar \omega\sum_{i=1}^{3}C_{ii}$ as an element.
The  $\lfr{rot}(3)$ algebra is spanned by three angular momentum
operators
\[ L_i = -i (C_{jk} - C_{kj}) \, ,\quad (i,j,k \; {\rm cyclic})\, ,\]
and five components $\{ Q_\mu; \mu = 0, \pm 1,\pm 2\}$ of the $L=2$
quadrupole tensor $Q$.

The Cartesian components of the monopole-quadrupole tensor are
expressed
\begin{equation}
{\Bbb Q}_{ij}=b_{0}^{2}\left[ A_{ij}+\textstyle\frac{1}{2}(C_{ij}+C_{ji})+B_{ij}
\right] \, .  \label{eq:monopole-quadrupoleACB}
\end{equation}
The spherical components of the quadrupole
operators are
\begin{equation}
Q_{\mu }=b_{0}^{2}\sqrt{3}(A_{2\mu }+C_{2\mu }+B_{2\mu })\,.
\label{eq:quadrupoleACB}
\end{equation}
One sees that the part of $Q_\mu$ that commutes with $H_0$ is the
$\lfr{su}(3)$ quadrupole operator
\[ {\cal Q}_\mu =  b_{0}^{2}\sqrt{3}C_{2\mu }\,. \]
Thus, $\lfr{su}(3)$ can be viewed as the projection of the $\lfr{rot}(3)$
algebra onto the space of operators that leave spherical harmonic
oscillator shells invariant.

An irrep of $\lfr{u}(3)$ is characterized by a highest weight state
$|\phi\rangle$ and a corresponding
highest weight $\omega = (\omega_1,\omega_2,\omega_3)$
defined such that
\[ C_{ij} |\phi\rangle = 0 \text{ for } i<j\,,\quad
C_{ii}|\phi\rangle = \omega_i|\phi\rangle \, .\]
Such an irrep remains irreducible on restriction to its $\lfr{u}(1) +
\lfr{su}(3) \subset \lfr{u}(3)$ subalgebra and has
$\lfr{u}(1) + \lfr{su}(3)$ highest weight $N(\lambda\mu)$, where
\[ N= \omega_1+\omega_2+\omega_3\,,\quad \lambda =
\omega_1-\omega_2\,,\quad \mu = \omega_2-\omega_3\,.\]

An irrep of the rigid-rotor
algebra, is characterized by an intrinsic state which is an eigenstate of
the quadrupole-moment operators with eigenvalues $\{ \bar Q_\mu \}$ that
are related to the shape variables of the Bohr-Mottelson collective
model \cite{BohMot52} by
\begin{eqnarray}
\overline{Q}_{0} &=&\sqrt{\frac{9}{5\pi }}AR_{0}^{2}\beta \cos \gamma \,,
\nonumber \\
\overline{Q}_{\pm 2} &=&\sqrt{\frac{9}{10\pi }}AR_{0}^{2}
\beta \sin \gamma \,,  \label{eq:betagamma} \\
\overline{Q}_{\pm 1} &=&0\,,  \nonumber
\end{eqnarray}
where $A$ is the mass number, $R_{0}$ is a nuclear radius, and $\beta $
and $\gamma $ are deformation and asymmetry parameters, respectively.

It is useful to note that the elements of $\lfr{sp}_{{\Bbb C}}(3,{\Bbb
R})$  are all compnents of U(3)\ tensors, i.e., they transform according
to irreducible representations of $\lfr{u}(3)$; the operators $\{A_{ij}\}$
are components of an irreducible tensor $A$ of highest weight $\{200\}$,
the $\{B_{ij}\}$ are components of an irrreducible tensor $B$
of highest weight $\{00-2\}$,  the operators
$\{C_{ij}, C_{ji},C_{ii}-C_{jj}; i\leq j\}$ are components of
an irreducible $\lfr{su}(3)$ tensor $C^{(11)}$ of $\lfr{su}(3)$ highest
weight $(11)$ ($\lfr{u}(3)$ highest weight $\{10-1\}$),  and $H_{0}=\hbar
\omega\sum_{i=1}^{3}C_{ii}$ is a u(3) scalar;
a tensor of highest weight $\{000\}$.

Now observe  that the first term, $H_{0}$, of the Hamiltonian
of Eq.\ (\ref{eq:Hsp3r}) is SU(3) (and U(3)) invariant whereas the second
term is invariant under the dynamical group ROT(3) $(= [{\Bbb
R}^{5}]$SO(3)) of a rigid rotor.
Thus, the two components, $H_0$ and $V(Q)$,  of the
Hamiltonian  respectively reduce
the two subalgebra chains:
\[
\begin{array}{ccc}
 & \lfr{sp}(3,{\Bbb R}) &  \\
\quad\swarrow &  & \searrow \quad\\
\lfr{su}(3)\quad &  & \quad\lfr{rot}(3) \\
\quad\searrow &  & \swarrow\quad \\
& \lfr{so}(3) &
\end{array}
\]
This implies that the eigenstates of $H = H_0 + V(Q)$ are
intermediate between those of the SU(3) and rigid-rotor models.
We show in the following that, in fact, both SU(3) and ROT(3) are
remarkable good quasi-dynamical symmetries for this Hamiltonian.

\section{Basis states and matrix elements}\label{sect:irreps}

A unitary irrep of $\lfr{sp}(3,{\Bbb R})$, within the shell-model space of
 a mass-$A$ nucleus, is characterized by a lowest-weight state
$|\sigma\text{lw}\rangle$, with weight $\sigma =
(\sigma_1,\sigma_2,\sigma_3)$, defined by the equations
\begin{equation}
B_{ij}|\sigma \text{lw}\rangle =0\,,\quad C_{ij}|\sigma \text{lw}\rangle =0%
\text{ for }i<j,\quad C_{ii}|\sigma \text{lw}\rangle =\sigma _{i}|\sigma
\text{lw}\rangle \,.  \label{eq:lw}
\end{equation}
When $\sigma$ is a triple of positive
integers or of positive half-odd integers, the corresponding irrep is 
either a discrete series representation or, for mass number $A<6$,
a limit of a discrete series irrep.

Let $\{ |\sigma \alpha\rangle\}$ denote an orthonormal basis for the
subspace of states of an $\lfr{sp}(3,{\Bbb R})$ irrep, of lowest weight
$\sigma$, that satisfy the equation
\beq B_{ij}|\sigma\alpha\rangle =0 \, .\eeq
 We refer to these states as {\it vacuum states\/} for the
corresponding   $\lfr{sp}(3,{\Bbb R})$ irrep.
They are a basis for a $\lfr{u}(3)$ irrep of highest
weight $\sigma$.
Moreover, it is known that a basis for an  $\lfr{sp}(3,{\Bbb R})$ irrep
can be constructed by acting on the vacuum  states with polynomials in the
$\{ A_{ij}\}$ raising operators.
Let
\beq Z^{(n)}_{KLM}(A) = [A\times A\times \ldots \times A]^{(n)}_{KLM} \eeq
denote a product of $N$ symplectic raising operators coupled to the
$(KLM)$ component of a U(3) tensor operator $Z^{(n)}(A)$ of highest weight
$n=(n_1,n_2,n_3)$, where $n_1$, $n_2$, and $n_3$ run over the even-integer
values for which
\beq n_1 \geq n_2\geq n_2 \geq 0 \, , \quad n_1+n_2+n_3 = 2N \, .\eeq
Acting on the vacuum states with these tensor operators gives  basis
states
\beq |\sigma n\rho\omega KLM\rangle = [Z^{(n)}(A) \times
|\sigma\rangle]^{\rho\omega}_{KLM} \label{eq:basis}\eeq
 which reduce the subgroup chain
\beq \matrix{{\rm Sp}(3,{\Bbb R}) &\supset& {\rm U}(3) &\supset& {\rm
SO(3)}&\supset &{\rm SO(2)} \cr
\sigma & n\rho& \omega &K&L&&M\cr} .\eeq
These states are eigenstates of the harmonic oscillator hamiltonian
\beq H_0  |(\sigma n)\rho\omega KLM\rangle =
(N_{0} + n_1+n_2+n_3)\hb\omega
 |(\sigma n)\rho\omega KLM\rangle \, ,\eeq
where
$N_0 = \sigma_1+\sigma_2+\sigma_3$.
However,  although the basis states of Eq.\ (\ref{eq:basis}) span
the carrier space of an $\lfr{sp}(3,{\Bbb R})$ irrep, they are not an
orthonormal basis due to a multiplicity of U(3) subirreps labelled by
the indices $n$ and $\rho$.
Thus, it is necessary to take suitable linear combinations
\[ |\sigma \tau \omega KLM\rangle = \sum_{n\rho}
(K^\sigma_{\omega})^{-1}_{n\rho,\tau} |\sigma n\rho\omega KLM\rangle
\]
 to form an orthonormal basis.
The K matrix coefficients are conveniently determined by vector coherent
state (VCS) methods \cite {Row84,Row85}.
The calculation of matrix elements of elements of the $\lfr{sp}(3,{\Bbb
R})$ Lie algebra in the corresponding orthonormal basis is also
straightforward using VCS theory.

Matrix elements of the components $\{X_{klm}\}$ of an SU(3) tensor
$X^{(pq)}$  are obtained from their $\lfr{su}(3)$-reduced matrix elements
using the generalized Wigner-Eckart theorem
\begin{eqnarray}
\langle \sigma \nu'\tau'\omega' K'L'M'|X_{klm}|\sigma \tau
\omega KLM\rangle  &=&\sum_{\gamma }((\lambda \mu )KL;(pq)kl\Vert \gamma
(\lambda'\mu')K'L')\,
(LM;lm|L'M')  \nonumber
\\
&&\times \langle \sigma \tau'\omega'\|
X^{(pq)}\| \sigma\tau \omega \rangle _{\gamma }\,,
\label{eq:Wigner-Eckart}
\end{eqnarray}
where $\lambda = \omega_1-\omega_2$, $\mu = \omega_2-\omega_3$, and
$\gamma $ indexes the multiplicity of SU(3) irreps of
highest weight $(\lambda'\mu')$ in the tensor product $(\lambda \mu
)\times (pq)$;
$(LM;lm|L'M')$ is an SO(3) Clebsch-Gordan coefficient and
$((\lambda \mu )KL;(pq)kl\| \gamma (\lambda'\mu' )K'L')$
is an  SO(3)-reduced Clebsch-Gordan coefficient for SU(3).

SU(3)-reduced matrix elements of elements of the  $\lfr{sp}(3,{\Bbb R})$
Lie algebra are given, for example, in ref.\ \cite{Row85}.
To calculate matrix elements of $Q\cdot Q$, following Rosensteel
\cite{Ros80},  we start by normal ordering the expansion
\begin{eqnarray}
Q\cdot Q &=&3(A_{2}+C_{2}+B_{2})\cdot (A_{2}+C_{2}+B_{2})  \nonumber \\
&=&6{\cal C}_{2}^{\lfr{su}(3)}-3L^{2}+10H_{0}+10\sqrt{6}B_{0}+6A_{2}\cdot
B_{2}+\{6C_{2}\cdot B_{2}+3B_{2}\cdot B_{2}+\text{h.c.}\}\,,
\label{eq:so3QQ}
\end{eqnarray}
using the commutation relations
\begin{eqnarray}
B_{2}\cdot A_{2}-A_{2}\cdot B_{2} &=&\textstyle\frac{10}{3}H_{0}\,, \nonumber \\
B_{2}\cdot C_{2}-C_{2}\cdot B_{2} &=&\textstyle\frac{10}{3}\sqrt{6}B_{0}\,.
\label{eq:comm}
\end{eqnarray}
With  such a normal-ordered expansion, i.e., with $B_2$ operators on the
right and $A_2$ operators on the left,  we do not have to
include intermediate states external to the truncated space;
consequently, the intermediate sums are minimized and the calculations are
less time consuming.
Optimization of the computations in this way is important because the
number of basis states grows exponentially as the number of major
oscillator shells in the calculation increases.
For example, suppose we want to calculate
matrix elements of $Q\cdot Q$ between states of a truncated space
comprising the oscillator shells
$N_{0}\hbar \omega $, $(N_{0}+2)\hbar \omega $, $(N_{0}+4)\hbar \omega $,
$\ldots $, $(N_{0}+n)\hbar \omega $.
Without normal ordering, we would have to
include intermediate states in the  calculation from the
$(N_{0}+n+2)\hbar\omega $ shell.
The number of extra states involved in the calculation would then be
proportional to $n(n+1)$.

The first three terms in Eq.\ (\ref{eq:so3QQ}) are diagonal in the chosen
basis with eigenvalues given by
\begin{eqnarray}
&\left\langle {\cal C}_{2}^{\lfr{su}(3)}\right\rangle =
\mathop{\textstyle\sum}
\nolimits_{i}(\omega _{i}-\textstyle\frac{1}{3}N)(\omega _{i}
-\textstyle\frac{1}{3}N-2i) = \textstyle\frac{2}{3}(
\lambda^2+\mu^2 + \lambda\mu + 3\lambda +3\mu) \,,&  \nonumber \\
&\left\langle L^{2}\right\rangle = L(L+1)\,, \quad \quad
\left\langle H_{0}\right\rangle = N\,. & \label{eq:u3ev} 
\end{eqnarray}
The quadratic terms are expressible as  $\lfr{su}(3)$-coupled tensors
using the identity
\begin{equation}
T_{2}^{(\lambda _{t}\mu _{t})}\cdot U_{2}^{(\lambda _{u}\mu _{u})}=
\sqrt{5}
\sum_{\rho (\lambda \mu )}\left\langle (\lambda _{t}\mu _{t})2;(\lambda
_{u}\mu _{u})2\Vert (\lambda \mu )0\right\rangle _{\rho }\left[ T^{(\lambda
_{t}\mu _{t})}\times U^{(\lambda _{u}\mu _{u})}\right]^{\rho (\lambda \mu
)}_0\,;  \label{eq:su3coup}
\end{equation}
this gives
\begin{eqnarray}
A_{2}\cdot B_{2} &=&\textstyle\frac{5}{6}\sqrt{6}(A\times
B)^{(00)}+\textstyle\frac{1}{6}\sqrt{30}(A\times B)^{(22)}_0\,,  \nonumber \\
C_{2}\cdot B_{2} &=&\sqrt{5}(C\times B)^{(02)}_0\,,  \label{eq:su3coupACB}
\\
 B_{2}\cdot B_{2} &=&\textstyle\frac{2}{3}\sqrt{5}(B\times B)^{(04)}_0 
+\textstyle\frac{5}{3}(B\times B)^{(20)}_0\,.  \nonumber
\end{eqnarray}
The $\lfr{su}(3)$-scalar component of $A_{2}\cdot B_{2}$ is related to the
quadratic Casimir invariants of $\lfr{sp}(3,{\Bbb R})$ and
$\lfr{su}(3)$
\[
(A\times B)^{(00)}=\textstyle\frac{1}{12}\sqrt{6}({\cal C}_{2}^{\lfr{su}(3)}
+\textstyle\frac{1}{3}H_{0}^{2}-4H_{0}-{\cal C}_{2}^{\lfr{sp}(3,{\Bbb R})})\,,
\]
where
\beq
\left\langle {\cal C}_{2}^{\lfr{sp}(3,{\Bbb R})}\right\rangle =
\mathop{\textstyle\sum}%
\nolimits_{i}\sigma _{i}(\sigma _{i}-2i)
= \textstyle\frac{2}{3}
(\lambda _{0}^{2}+\lambda _{0}\mu _{0}+\mu _{0}^{2}+3\lambda _{0}+
3\mu_{0})+\textstyle\frac{1}{3}N_{0}^{2}-4N_{0} \,,\label{eq:c2sp3} \\
\eeq
with
\beq \lambda_0 = \sigma_1-\sigma_2\,,\quad \mu_0 =
\sigma_2-\sigma_3\,,\quad N_0 =\sigma_1+\sigma_2+\sigma_3\,.
\eeq

Combining the above results, Eq. (\ref{eq:so3QQ}) becomes
\begin{eqnarray}
Q\cdot Q &=&-\textstyle\frac{5}{2}{\cal C}_{2}^{\lfr{sp}(3,{\Bbb R})}
+\textstyle\frac{17}{2}{\cal C}_{2}^{\lfr{su}(3)}-3L^{2}
+\textstyle\frac{5}{6}H_{0}+\sqrt{30}(A\times B)^{(22)}_0
\nonumber \\
&&+\{10\sqrt{6}B_{0}+6\sqrt{5}(C\times B)^{(02)}_0  \nonumber \\
&&+2\sqrt{5}(B\times B)^{(04)}_0+5(B\times B)^{(20)}_0+\text{h.c.}\}\,.
\label{eq:su3QQ}
\end{eqnarray}

\section{Results}

\label{sect:results} Calculations have been done for a typical heavy
nucleus with oscillator energy $\hbar \omega =7.49$ MeV; a value
appropriate for the $_{\; 68}^{166}$Er nucleus.
The deformation parameter $\varepsilon =1.01\times 10^{11}$
gives the minimum of the collective potential at
$\beta _{{\rm in}}=0.35$; a value close to that inferred from the
experimental $B(E2$:$2_{1}^{+}\rightarrow 0_{1}^{+})$ transition
rate.  We choose an $\lfr{sp}(3,{\Bbb R})$ irrep with lowest
weight  $(327\frac{1}{2},249\frac{1}{2},249\frac{1}{2})$
(or equivalently, $826\frac{1}{2}(78,0)$ in a U(1) $\times$ SU(3)
notation \cite{RosRow80}).
This irrep is deduced from an empirical
formula for the intrinsic mass quadrupole moment of a deformed oscillator
\cite{Jarrio,Row85}.
The strength $\chi$ of the potential $V(Q)$ in Eq.
(\ref{eq:V_Q}) is  varied in this study.
 The ratio between the potential and oscillator strengths determines the
structure of the wave functions. Due to computational limitations, the
space is restricted to states belonging to shells below $12\hbar \omega$;
apart from this restriction there is no further truncation.

\subsection{Ground state band}

\begin{figure}[tbp]
\epsfxsize=6in
\centerline{\epsfbox{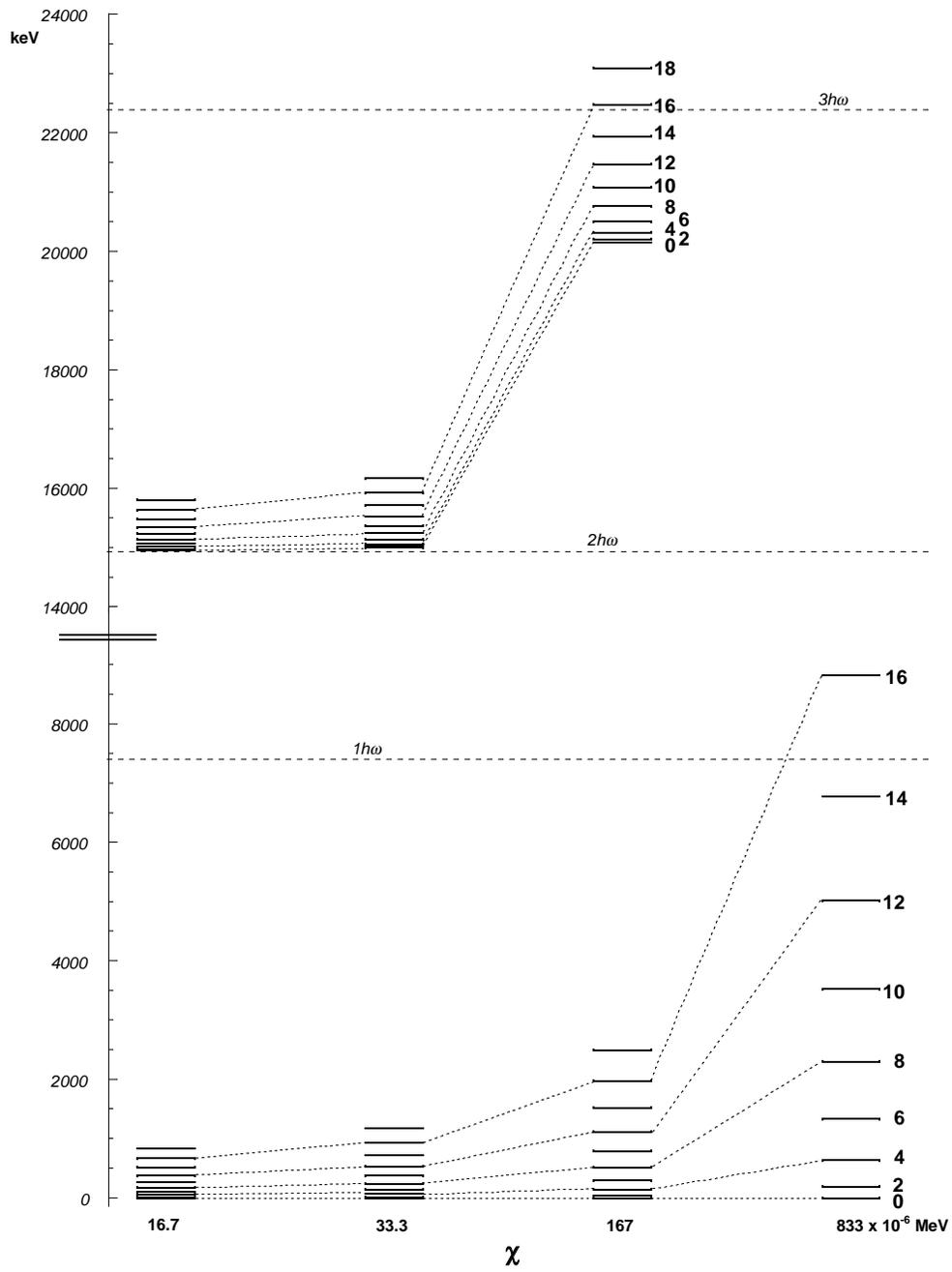}}
\caption{The energy spectra for various strengths of the potential. The
dashed lines connect energy levels
of the same
angular momenta.}
\label{fig:spectra}
\end{figure}

Energy spectra for different $\chi$ values (in units of MeV) are given in
Fig.\ \ref{fig:spectra}.
The results show ground state bands with
rotational spectra and  moments of inertia that decrease
as $\chi$ increases.
Moments of inertia, defined for each angular momentum state, by the
expression
\begin{equation}
{\cal J}_L = \frac{1}{2(E_L-E_0)} L(L+1)
\end{equation}
are shown in Table \ref{table:momentinertia}.

\begin{table}[th]
\caption{Moments of inertia for states of angular momentum $L$; the
values in parenthesis are percentages relative to the value at $L=2$.}
\label{table:momentinertia}
\begin{tabular}{ccccc}
$L\;\backslash \;\chi $ & 1.67$\times $10$^{-5}$ MeV & 3.33$\times $10$^{-5}$
MeV & 1.67$\times $10$^{-4}$ MeV & 8.33$\times $10$^{-4}$ MeV \\ \hline
2 & 207.62 MeV$^{-1}$ (100.00\%) & 148.27 MeV$^{-1}$ (100.00\%) & 70.10 MeV$%
^{-1}$ (100.00\%) & 15.72 MeV$^{-1}$ (100.00\%) \\
4 & 207.44 (99.91\%) & 148.15 (99.92\%) & 70.04 (99.91\%) & 15.70 (99.89\%)
\\
6 & 207.16 (99.78\%) & 147.97 (99.80\%) & 69.94 (99.76\%) & 15.68 (99.72\%)
\\
8 & 206.78 (99.59\%) & 147.72 (99.63\%) & 69.79 (99.55\%) & 15.64 (99.49\%)
\\
10 & 206.29 (99.36\%) & 147.41 (99.42\%) & 69.61 (99.30\%) & 15.59 (99.20\%)
\\
12 & 205.71 (99.08\%) & 147.02 (99.16\%) & 69.40 (98.99\%) & 15.54 (98.85\%)
\\
14 & 205.02 (98.75\%) & 146.57 (98.86\%) & 69.14 (98.62\%) & 15.47 (98.43\%)
\\
16 & 204.22 (98.36\%) & 146.06 (98.51\%) & 68.84 (98.20\%) & 15.40 (97.95\%)
\\
18 & 203.33 (97.93\%) & 145.48 (98.12\%) & 68.50 (97.72\%) & 15.31 (97.41\%)
\end{tabular}
\end{table}

The most remarkable result is that, for each of the interaction strengths
shown, the spectra are almost identical to those of a rigid rotor with
excitation energies very accurately proportional to $L(L+1)$.
$B(E2)$ transition rates between adjacent states, shown for $\chi =
3.33\times 10^{-4}$ MeV in Table \ref{table:be2}, are also in remarkably
close agreement with those of a rigid rotor.
These results are significant because, they give rotor model results with
fully microscopic 166-particle wave functions.
Thus, they provide us with the means to explore the rotational dynamics
of a nucleus at the microscopic, many-nucleon, level.  As a comparison,
the results from Elliott's SU(3) model are also given with the effective
charge $e=2.04$.  In the symplectic-Davidson model, $e=1$.

\begin{table}[th]
\caption{The reduced quadrupole transition strength
$B(E2:L\rightarrow L{-}2)$ in Weisskopf unit (W.u.)
for the experimental data, SU(3) model, this model (with coupling 
constant $\chi=3.33\times 10^{-4}$ MeV), and a rigid rotor. Ratios of
transition strengths to those of the the
$2\to 0$ transition are given in parenthesis.}
\label{table:be2}
\begin{tabular}{ccccc}
$L_i \rightarrow L_f$ & experiment & SU(3) & Davidson & rigid rotor \\ 
\hline
 $2 \rightarrow  0$ & $214 \pm 10$ (1.00) & 207 (1.00) & 207 (1.00) & 207 (1.00) \\
 $4 \rightarrow  2$ & $311 \pm 20$ (1.45) & 295 (1.43) & 295 (1.43) & 296 (1.43)  \\
 $6 \rightarrow  4$ & $347 \pm 45$ (1.62) & 324 (1.57) & 325 (1.57) & 326 (1.57)  \\
 $8 \rightarrow  6$ & $365 \pm 50$ (1.70) & 338 (1.63) & 339 (1.64) & 340 (1.64)  \\
$10 \rightarrow  8$ & $371 \pm 46$ (1.73) & 345 (1.68) & 348 (1.68) & 350 (1.69)  \\
$12 \rightarrow 10$ & $376 \pm 40$ (1.76) & 349 (1.69) & 353 (1.71) & 356 (1.72)  \\
$14 \rightarrow 12$ & & 350 (1.69) & 356 (1.72) & 361 (1.74) \\ 
$16 \rightarrow 14$ & & 350 (1.69) & 357 (1.73) & 365 (1.76) \\
$18 \rightarrow 16$ & & 349 (1.69) & 358 (1.73) & 367 (1.77)
\end{tabular}
\end{table}

A question of considerable interest is the nature of nuclear rotational
energies.
In the Bohr-Mottelson rotor model \cite{BohMot52} rotational energies are
interpreted as arising from the kinetic energy whereas in Elliott's
SU(3) model \cite{Ell58} they come from the potential energy.
Since the symplectic model contains both a rigid-rotor and Elliott's
SU(3) models as limiting submodels, it is of considerable interest to see
how it interpolates between the two limits.
In the symplectic model, the kinetic energy
operator is written
\begin{equation}
T=\textstyle\frac{1}{2}\hbar \omega \left[ H_{0}-\sqrt{\textstyle{\frac{3}{2}%
}}(A_{0}+B_{0})\right] \,.
\end{equation}
The contribution of the kinetic to the total
calculated excitation energy is shown as a percentage for each state of
the ground state band in Table \ref{table:kinetic}).
\begin{table}[th]
\caption{The kinetic energy as
a percentage of the total excitation energy for different values of
the coupling constant
$\protect\chi $. The values of $\protect\beta _{0}$ show the average
deformation of the ($L=0$) ground state   according to Eq.\
(\protect\ref{eq:betagamma}). }
\label{table:kinetic}
\begin{tabular}{ccccc}
$L\;\backslash \;\chi $ & 1.67$\times $10$^{-5}$ MeV & 3.33$\times $10$^{-5}$ 
MeV & 1.67$\times $10$^{-4} MeV $ & 8.33$\times $10$^{-4}$ MeV \\ \hline
2 & -31.95\% & -21.57\% & 26.43\% & 9.46\% \\
4 & -31.93 & -21.56 & 26.44 & 9.46 \\
6 & -31.90 & -21.53 & 26.46 & 9.45 \\
8 & -31.85 & -21.49 & 26.48 & 9.44 \\
10 & -31.79 & -21.44 & 26.52 & 9.43 \\
12 & -31.72 & -21.37 & 26.55 & 9.41 \\
14 & -31.64 & -21.30 & 26.60 & 9.39 \\
16 & -31.54 & -21.22 & 26.65 & 9.37 \\
18 & -31.43 & -21.12 & 26.71 & 9.35 \\ \hline
$\beta_0$ & 0.247 & 0.275 & 0.345 & 0.351
\end{tabular}
\end{table}
It can be seen that, for the smaller values of $\chi$, the
kinetic energy gives a negative contribution to excitation energies.
Its contribution is positive for larger values of
$\chi$, but remains much less than that of the potential energy.
Note, however, that because of the truncation to
shells of $12\hbar\omega$ and below, the results shown for $\chi =
8.33\times 10^{-4}$ MeV
are not fully converged and are unreliable.
Even so, since we obtain rotational bands in close agreement with the
rotor model for a wide range of potential strengths, it is hard to
avoid the inference that nuclear rotational energies are most likely
not 100\% kinetic in origin.
If this  is correct, it has considerable conceptual implications for the
interpretation of nuclear rotational dynamics.
We may continue, for convenience, to describe the cofactor of the $L(L+1)$
rotational energy as the inverse of a ``moment of inertia", but one
should recognize that, if the excitation energy is not kinetic, the
concept is misleading.

\begin{figure}[tbp]
\epsfxsize=6.5in
\centerline{\epsfbox{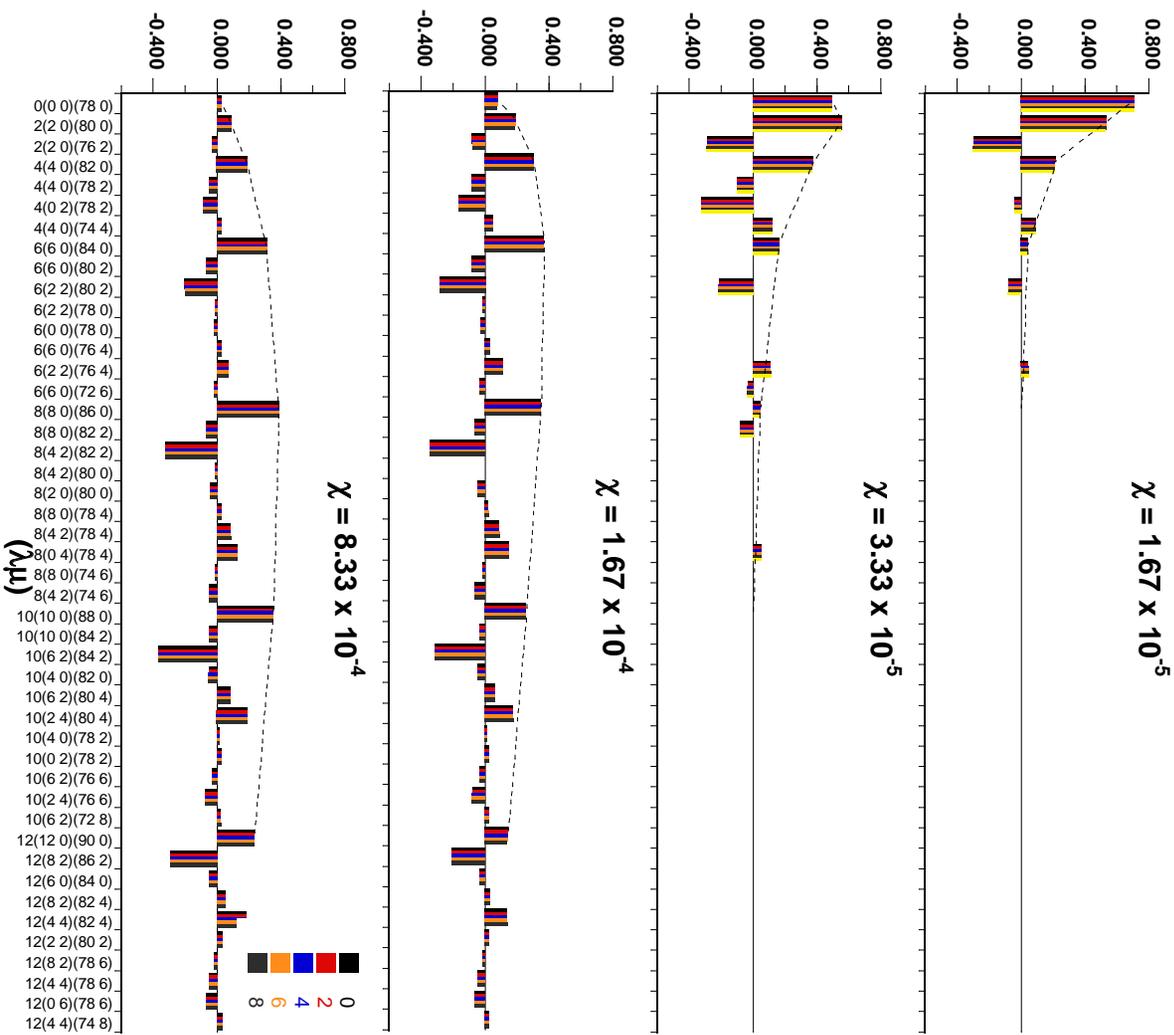}}
\caption{Histograms of the SU(3) amplitudes for the yrast states with
variation of the strengths of the potential. The
so-called  stretched states are connected with dashed
lines.}
\label{fig:Dvdev}
\end{figure}

Another result of this study is that the moments of inertia as
well as the kinetic energy portions of the excitation energy change less
than 5\% over the range of angular-momentum values considered.
This result can be interpreted as signifying that the states of a band
have a common intrinsic structure that changes little with increasing
angular momentum.
This interpretation becomes much more compelling when one observes the
behaviour of the coefficients of the wave functions in the expansion
\begin{equation}
| \psi_{K L M}\rangle = \displaystyle \sum_{(\lambda\mu)} |
{(\lambda\mu) K L M}\rangle C_{(\lambda\mu)KL} \, ,
\label{eq:adiabat}
\end{equation}
of the states of the ground state band in a U(3) $\supset$ SU(3)
$\supset$ SO(3) basis.
The coefficients are plotted for different $\chi$
values in Fig.\  \ref{fig:Dvdev}.
The figure shows clearly, that for all values of $\chi$ considered, the
coefficients are essentially independent of $L$ and negligible for
$K\not= 0$.
Moreover, for the larger $\chi$, this is in spite of a huge
mixing of SU(3) irreps from many major shells.
It follows that, while SU(3) is far from being a good dynamical symmetry,
it remains an extrordinarily good quasi-dynamical symmetry
according to the definition given in the Introduction and Refs.\
\cite{BahRowWij98,RowRoc88}.

Observe also that the distribution of U(3) irreps is dominated by
the so-called stretched irreps;  the stretched irreps are those of
the sequence
\[  N_0(\lambda_0,0) \, ,\quad N_0+2(\lambda_0+2,0)\,,\quad
N_0+4(\lambda_0+4,0)\,, \quad \ldots , \quad N_0+2n(\lambda_0+2n ,0) \, ,
\quad \ldots \, .\]

\begin{figure}[tbp]
\vspace{7in}
\includegraphics{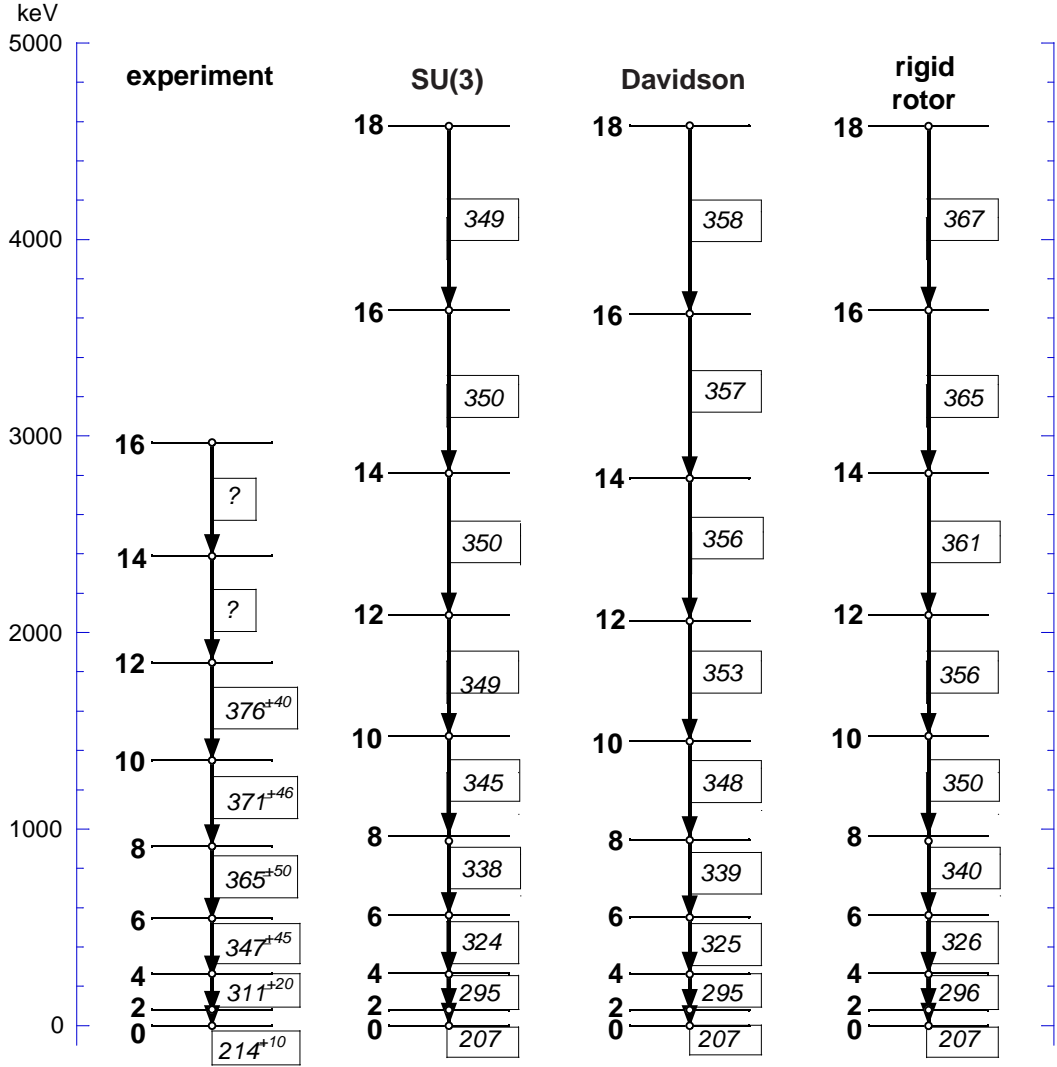}
\caption{Energy spectra for experimental data, the SU(3), the 
symplectic-Davidson, and the rigid rotor models.  The $B(E2)$ values are
given inside the boxes.}
\label{fig:energy}
\end{figure}

\begin{figure}[tbp]
\epsfxsize=6in
\centerline{\epsfbox{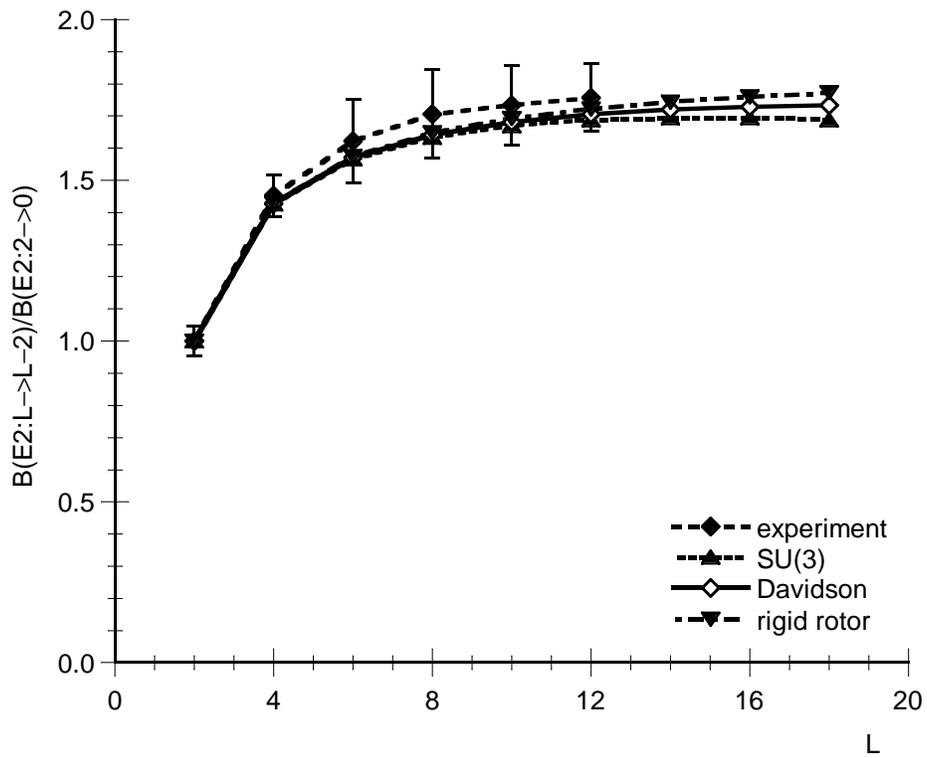}}
\caption{Relative $B(E2)$ values for experimental data, the SU(3),
the symplectic-Davidson, and the rigid rotor models.}
\label{fig:be2}
\end{figure}

The value of the coupling constant for which the $L=0$ ground state of
the model Hamiltonian has a deformation $\beta_{0}=0.350$ equal to
$\beta_{\rm in}$, the value for which the Davidson potential is a
minimum, was found by repeated calculation to be given by
$\chi=3.33\times 10^{-4}$ MeV. We call this the self-consistent coupling
constant. The spectrum of the ground band for this $\chi$ is displayed in
Fig.\ \ref {fig:energy} in comparison with the observed spectrum of
$^{166}$Er and that of the rigid rotor.
One sees that the results track the rigid rotor more closely than they do
experiment.
This may be an artifact of the Davidson potential which
continues to rise unrealistically at large deformation and excessively
inhibits centrifugal stretching.
 We also computed the  reduced electromagnetic
transition strengths
\begin{equation}
B(E2:L_{i}\rightarrow L_{f})={\frac{2L_{f}+1}{2L_{i}+1}}\left( {\frac{5}{%
16\pi }}\right) \left( {\frac{eZ}{A}}\right) ^{2}|\langle f||Q||i\rangle
|^{2}\,.  \label{eq:be2}
\end{equation}
The ratios of transition strengths, shown in parenthesis in Table
\ref{table:be2} and also depicted in Fig. \ref{fig:be2}, 
are in extremely close agreement both with experiment
and the rigid-rotor model.
Their magnitudes shown in   Fig.\ \ref{fig:energy}  are also in good
agreement with experiment.

\subsection{Giant resonance bands}

In addition to the ground-state rotational band, the symplectic
model gives many excited bands.
However, for a single $N_0(\lambda_0, 0)$ irrep, as considered here,
without spin-orbit, pairing and other
irrep-mixing interactions, the
simple symplectic model has no low-lying excited bands any more than
the SU(3) model  has excited bands for a $(\lambda_0,0)$ irrep. The
lowest-energy excited bands of the simple symplectic model are associated
with the giant monopole (breathing mode) and giant quadrupole resonance
degrees of freedom.

\begin{figure}[tbp]
\epsfxsize=6in
\centerline{\epsfbox{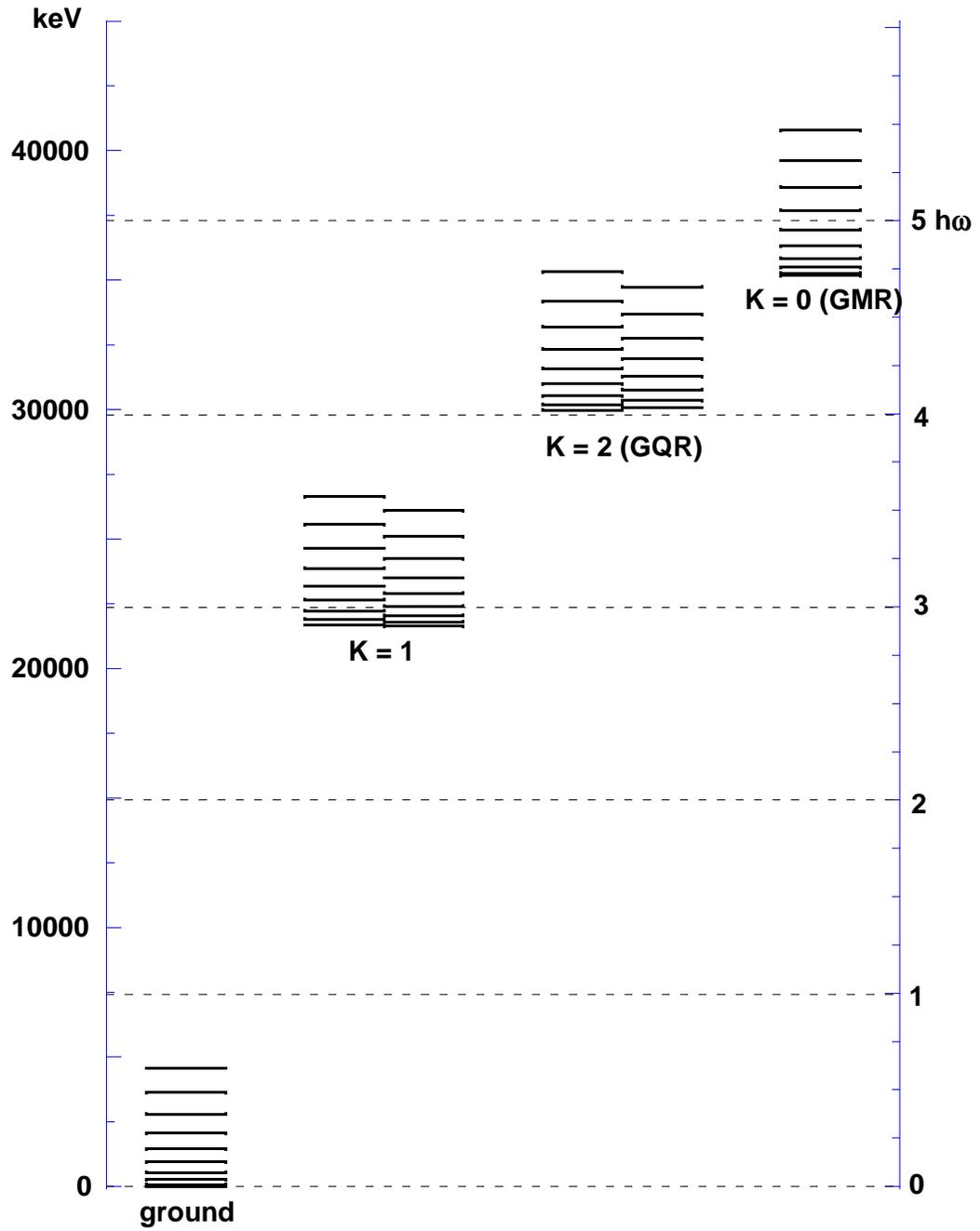}}
\caption{Energy spectra in the Davidson-symplectic model for the
ground-state and breathing-mode (giant-beta)  bands with the
self-consistent parameter strength $\chi=3.33\times 10^{-4}$ MeV.}
\label{fig:spectrum_self}
\end{figure}

For small values of $\chi$, these occur in the model, with a Davidson
potential, at around $2\hbar\omega$.
They are shown for several values of $\chi$ in Fig.\ \ref {fig:spectra};
results for the self-consistent value are given in 
Fig.\ \ref{fig:spectrum_self}.

Two results are worth noting.
The first is that the energies of the giant-resonance bands rise with
increasing values of $\chi$ and become unrealistically high at the value
considered appropriate for the ground state band.
This we believe to be a reflection of the fact that, although the Davidson
potential has many useful features, it rises too steeply away from the
equilibrium deformation.
The second notable result is that the symplectic model gives three
monopole-quadrupole giant resonance bands; two of these,  the $K=0$ and
$K=2$ bands, occur also (albeit at much lower energy) in the
Bohr-Mottelson model where they are associated with beta- and 
gamma-vibrations, respectively.
However, whereas there is no $K=1$ one-phonon band in the Bohr-Mottelson
model, such a band is non-spurious in the symplectic model which includes
intrinsic vorticity degrees of freedom.

\begin{figure}[tbp]
\epsfxsize=6in
\centerline{\epsfbox{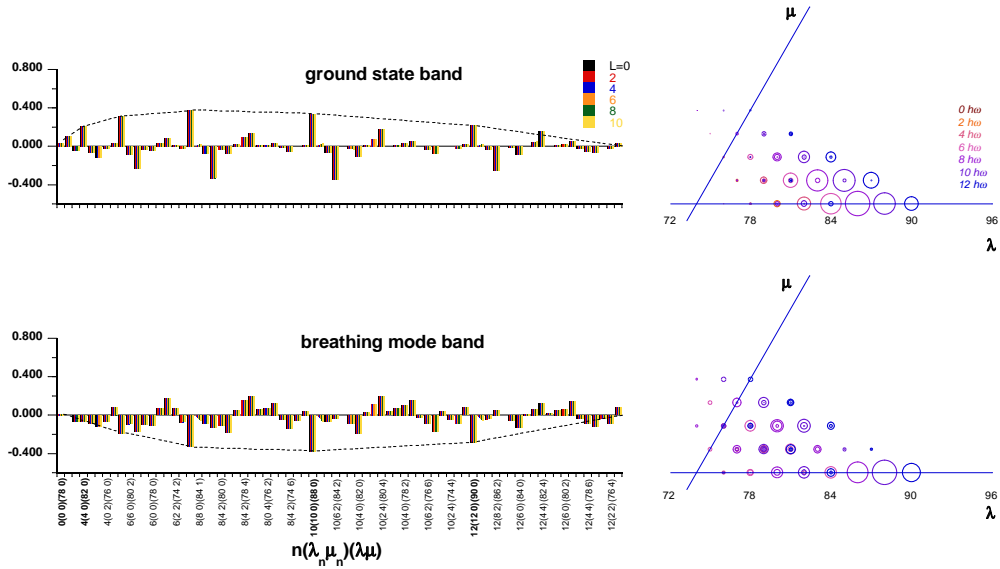}}
\caption{Histograms and corresonding plots on an SU(3) weight
diagram of the SU(3) amplitude coefficients for the
ground-state and breathing-mode (giant-beta) bands.}
\label{fig:diagram}
\end{figure}

\begin{figure}[tbp]
\epsfxsize=6in
\centerline{\epsfbox{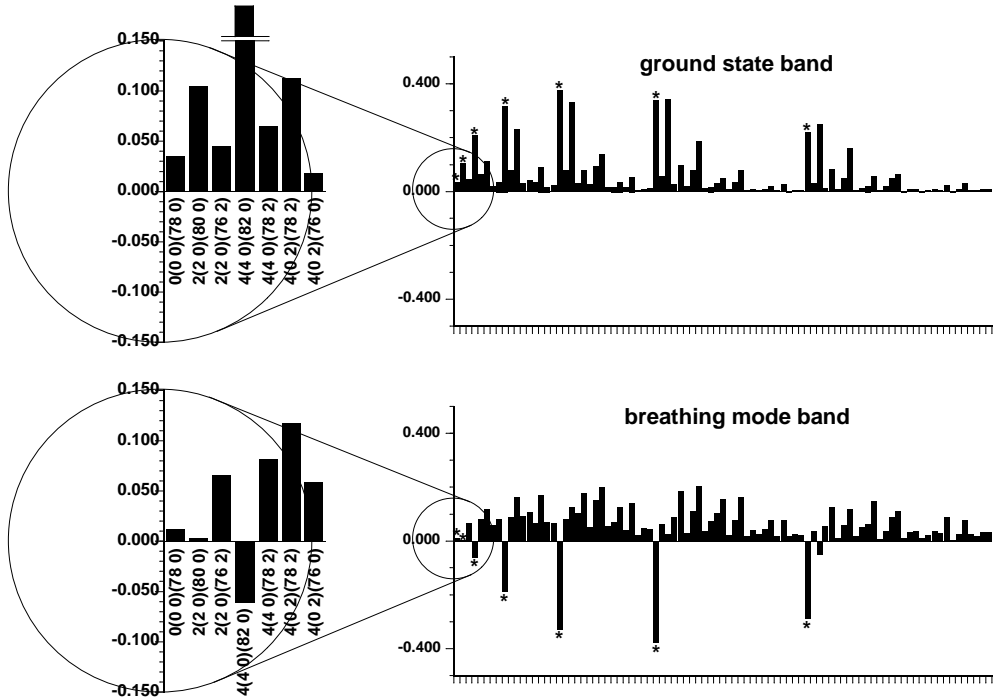}}
\caption{As for Fig.\ \protect \ref{fig:diagram} except that only
$L=0$ coefficients are plotted and the phase factor for the
ground-state coefficients are set to be
positive. The stretched bases are indicated by asterisks. The insets show
changes of phase for the first few coefficients.}
\label{fig:evabs}
\end{figure}

The monopole and quadrupole giant resonance states in real nuclei are
believed to lie at energies close to $2\hbar\omega$.
However, their strength is  invariably fragmented and much of it lies in
the continuum.
Thus, we make no attempt to compare our results for the GR states with
experiment.
However, it is of interest to examine the structure of the
wave functions.
Fig.\ \ref{fig:diagram} shows the decomposition of the
$K=0$ (giant-beta band) states in terms of their SU(3) components, for the
self-consistent value of $\chi$, in comparison with the corresponding
decomposition of the ground-band states.
It can be seen that the amplitude
coefficients (including their signs)  are independent of
angular momentum to a high degree of accuracy.
It can also be seen that the coefficients for the two sets of states have
the same signs except for those of the stretched states, which are of
opposite sign.
Note that the relative signs of the amplitude
coefficients for different SU(3) subirreps are determined by SU(3)
Clebsch-Gordan coeffients \cite{DraAki73} and have no particular meaning,
in general, except for the stretched states which all have the same sign.
However, a change in the relative signs of coefficients between the ground
and excited states is meaningful.
Thus, it is fortuitous that the coefficients for the stretched states all
have positive sign for the states of the ground state band because it
highlights the fact that these coefficients
change sign for the giant beta band.
On reflection this is what one would expect for a giant-beta vibrational
excitation (see Fig.\ \ref{fig:evabs}).

\section{Conclusion}\label{sect:conclusions}

The symplectic model is currently the only model that is capable of giving
rotational states for heavy nuclei as eigenstates of a
rotationally-invariant Hamiltonian in a realistic shell-model space.
Thus, the ability of
the model to give the energy levels and $E2$ transitions strengths between
states for ground-state rotational bands, without the use of an
effective charge, provides a powerful framework for
understanding the dynamics of nuclear rotations in terms of interacting
neutrons and protons.

Early applications of the model by Park {\it et al.} \cite{ParkAl84} were
remarkably successful and raised many interesting questions.
However, because of the severe truncations of the space that were
necessary at the time, the reliability of the results
could be questioned.
In particular, one was concerned that the ability of the model to give
correct moments of inertia might be lost on increasing the size of the
model space. The results of the present calculations show that this is
not the case.

A particularly interesting challenge was to learn how a model, without
pair correlations, could give correct moments of inertia when it is
known that the cranking model is only succesful when pairing correlations
are included.
The early calculations of Park {\it et al.} indicated that the dominant
contribution to rotational energies came from the potential energy part
of the Hamiltonian, thus calling into question the very concept of the
moment of inertia as an inverse coefficient of the $L^2$ term in the
kinetic energy.
The results of the present calculation indicate that the inclusion of only
stretched states, as in the calculation of Park {\it et al.}, tends to
exagerate this effect.
Nevertheless, it confirms that the dominant component of the rotational
energies comes from the potential energy; for the self-consistent value
of $\chi$ only about 20\% of the rotational energy
comes from the kinetic energy in the present calculation.

In addition to investigating rotational states up to much higher angular
momentum $(L=18)$, the present calculations have focussed on
understanding the structure of rotational states in terms of their SU(3)
content.
We have shown that although there is huge mixing of SU(3) irreps from
many major harmonic oscillator shells, the mixing is highly coherent and
establishes SU(3) as a remarkably good quasi-dynamical symmetry for the
model.
Other calculations \cite{RowRoc88,BahRowWij98}, reviewed in \cite{NAC},
show that such quasi-dynamical symmetry is also conserved when SU(3)
irreps are further mixed by spin-orbit and pairing forces.
Thus, the results show that, as far as the calculation of $E2$ transition
rates are concerned, the use of a single SU(3) irrep with an effective
charge will give accurate results.
However, for other observables, not related to elements in the SU(3)
algebra, there is no reason whatever to expect effective charge
methods to take account of the large mixing of SU(3) irreps observed.

The close agreement between the results of the symplectic model
calculation and those of the rigid-rotor model, shows that, by
definition, the rigid-rotor algebra is also an excellent quasi-dynamical
symmetry for the symplectic model with a Davidson potential.
But again, for observables not related to the rigid-rotor algebra, such
as electron scattering current operators, one cannot predict what the
results will be by purely algebraic methods.

The fact that two competing dynamical subgroup chains, although very
different in their physical content, can both be good quasi-dynamical
symmetries is remarkable but understandable.
It can happen because, for large-dimensional representations and states of
relatively low angular momentum, an SU(3) irrep contracts to an irrep of
the rigid-rotor algebra \cite{C84}.
Thus, it transpires that the lower angular-momentum states of a
large-dimensional SU(3) irrep belong to an embedded representation
\cite{RowRoc88} of the rigid-rotor algebra and vice-versa.

Having demonstrated that symplectic model calculations with
phenomenological (albeit microscopically expressable) potentials have the
ability to describe nuclear rotational states, our next goal would
be ideally to perform calculations within the same (large) shell model
space but with realistic two-nucleon interactions.
Such calculations can and have been contemplated for light nuclei
\cite{VCR}. But they are computer intensive and impractical for
heavy rotational nuclei.
We would even like to go further and include spin-orbit and
short-range (e.g., pairing)  interactions which mix different
$\lfr{sp}(3,\Bbb R)$ irreps.
We would like to carry out calculations for superdeformed
bands \cite{Twin86,BakHaaNaz95} as well as for normally deformed
low-lying rotational bands of heavy nuclei.
Superdeformed bands are naturally associated with
excited representations of the symplectic model which fall into the
low-energy domain (as do Nilsson model states) as a consequence
of shell effects in a deformed shell model.
The challenge is to explain why they do not mix more readily
with the large density of less deformed low-energy states.
We believe that symplectic symmetry and su(3) quasi-dynamical symmetry
has the potential to answer such questions.

It is unlikely that such calculations will ever be done in
a shell-model space sufficiently large to ensure convergence of the
results.
 However, the results of the present study and previous
investigations of mixing SU(3) irreps with spin-orbit \cite{RowRoc88} and
pairing  interactions\cite{BahRowWij98} imply that, when experiment
finds states of a nucleus that are fitted well by the
rotor model and to have a large deformation, then we
have reason to believe that SU(3) is a good quasi-dynamical symmetry
for these states.
Armed with this information, one can hope to design realistic mixed
SU(3) calculations within large spaces.
In particular,   one can calculate  just
one representative angular momentum state for each band, with the
understanding that the coefficients should be the same (to a good
approximation) for all other states of the band. Alternatively, one could
carry out calculations in an intrinsic space which includes just the
highest weight states for the contributing SU(3) irreps.

\acknowledgments

This investigation was supported by the Natural Sciences and Engineering
Research Council of Canada.
We thank the Institute for Nuclear Theory at the University of Washington
for its hospitality and the U.S. Department of Energy for partial support
during the final stage of this work.

\newpage

\end{document}